\documentstyle[12pt,epsf]{article}
\newcommand{\etal}{{\it et al.}}

\def\Journal#1#2#3#4{{#1} {\bf #2}, #3 (#4)}

\def\NIMA{{Nucl. Instrum. Methods} A}
\def\NPB{{Nucl. Phys.} B}
\def\PLB{{Phys. Lett.}  B}
\def\PRL{Phys. Rev. Lett.}
\def\PRD{{Phys. Rev.} D}
\def\ZPC{{Z. Phys.} C}
\def\NPV{{Nucl. Phys.} V}

\begin{document}

\title{\boldmath Prospects for Searching for Excited
Leptons during Run~II of the Fermilab Tevatron}

\maketitle


\begin{centering}
{
  E. Boos$^1$,  A. Vologdin$^1$, D. Toback$^2$, and J. Gaspard$^2$\\
  \hfil\\
  $^1$Institute of Nuclear Physics, Moscow State University,
      119899, Moscow, Russia \\
  $^2$Texas A\&M University, College Station, TX 77843  USA}

\end{centering}

\begin{abstract}
This letter presents a study of prospects of searching for excited
leptons during Run~II of the Fermilab Tevatron.  We concentrate on
single and pair production of excited electrons in
the photonic decay channel in one CDF/D$\O$ detector equivalent for 
2 fb$^{-1}$.  By the end of Run~IIa, the limits should be easily extended 
beyond those set by LEP and HERA for excited leptons with mass above about 
190 GeV.

\end{abstract}


\clearpage


The Standard Model (SM) of particle physics is known to give results that 
match the current experimental data with a high precision.
However, because of well known theoretical problems and disadvantages,
it is widely believed it can not be a complete theory of elementary 
particles, but rather, likely a kind of effective theory at energies 
below some scale of the order of a TeV.  While many models of ``new physics" 
beyond the SM have been suggested over the years, one of the most 
straight-forward ideas proposes that quarks and leptons are composite 
particles.
Such a scenario is a recurring theme in nature;  molecules are
composed of  atoms, which are in turn composed of nuclei, which are in
turn composed of  nucleons and so on.  Furthermore, composite models 
can explain, in principle, family replication, mixing in the quark 
and lepton sectors, and even make the fermion masses and weak mixing 
angles calculable.

In most composite models fermions possess an
underlying substructure which is characterized by a
scale $\Lambda$~\cite{Other} with  $\Lambda$ about 1 TeV or higher~\cite{g-2}.
While there is no unified model of compositeness, a model-independent
effective Lagrangian for excited leptons, originally proposed
in~\cite{Hagiwara}, can be used to model single and pair production in experiments.  The Lagrangian:

\begin{eqnarray}\label{EffLagrangian}
L_{l^{*}l^{*}}  &=&  \overline{l}^{*}\gamma^{\mu}(\frac{\tau}{2}W_{\mu}+g^{'}\frac{Y}{2}B_
{\mu})l^{*}\\
L_{l^{*}l}  &=&  \frac{1}{2\Lambda} \overline{l}^{*}_{R} \sigma^{\mu\nu}
(fg\frac{\tau}{2}W_{\mu\nu} + f^{'}g^{'}\frac{Y}{2}B_{\mu\nu} ) l_{L}
+h.c.
\end{eqnarray}

\noindent
where $f$ and $f'$ are coupling constants, has been used extensively in a number of phenomenological papers which 
presents ideas on searching for excited fermion production and decay to 
final state gauge bosons in $e^+e^-$, $p\bar{p}$, $ep$, and $e \gamma$ 
collisions~\cite{Boudjema2,Baur,Boos}.  Direct searches for lepton 
compositeness has been done extensively at  LEP~\cite{LEP} and 
HERA~\cite{HERA}, each with no discovery, but with ever more sensitive limits. 
Unfortunately, only direct searches for quark compositeness have 
been done at the Fermilab Tevatron~\cite{Tevatron Quark}.  





In this Letter we present a study for searching for excited lepton production
and decay for the upcoming Run~II of CDF and D$\O$ detectors.
We begin with an updated simulation of excited lepton production, and optimize
the sensitivity by studying the kinematics of excited lepton production and 
decay.  With these results we estimate the mass reach and compare to the 
recent results from LEP and HERA. 




To study the mass reach of the Tevatron, we simulated single and pair
production and decay of excited leptons using the upgraded Fermilab accelerator
(1.8 $\rightarrow$ 2.0 TeV), and CDF and D$\O$ detectors for Run~II.  The 
Feynman rules from the effective Lagrangian (Eqn.~\ref{EffLagrangian}) are
implemented into {\sc CompHEP}~\cite{CompHEP} using the {\sc LanHEP}~\cite{LanHEP} 
software package.
We have included into this simulation a complete tree-level calculation
which takes into account all the spin correlations between excited
states production, subsequent decays, and incorporate the known NLO 
corrections.  All the partial width and known 2$ \rightarrow$ 2 
cross-section have been cross-checked at the symbolic level.  Events 
on parton level generated by means of {\sc CompHEP} have been used as an 
external process for {\sc pythia} with the help of the {\sc CompHEP}-{\sc pythia} 
interface \cite{interface}.  The underlying event, jet fragmentation, 
ISR and FSR are modeled using the {\sc pythia}~\cite{Pythia} Monte Carlo 
with the {\sc cteq4l}~\cite{CTEQ4L} structure functions.
Since Drell-Yan production of excited leptons production 
is similar to that of Supersymmetric leptons, we take the K-factors 
 given in \cite{K-signal}, which only depend on  the  masses of 
final particles, and vary between 1.23 and 1.24 in the mass range 
of the search.

\begin{figure}[htc]
  \vspace*{-.75in}
  \epsfbox{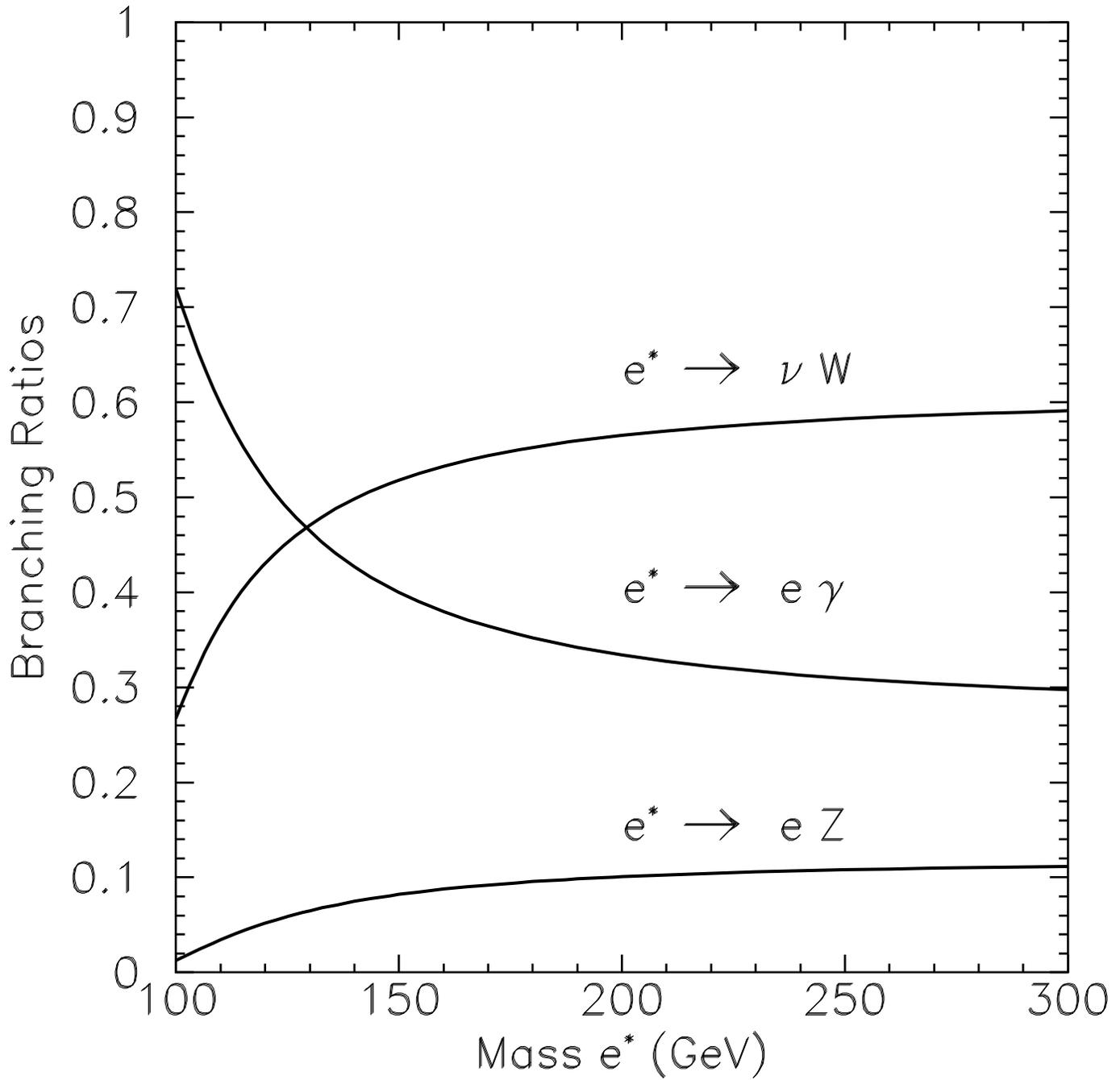}
  \vspace*{-1in}
  \caption{The branching ratios for excited leptons as a function of the mass of the excited lepton.}
  \label{BranchingRatios}
  \end{figure}

While excited leptons can come in three flavors: $e^{*}$, $\mu^{*}$ and $\tau^{*}$, 
we chose to concentrate on the electron since the result of 
$\mu^{*}$ expected to be similar to $e^{*}$, and $\tau$'s at the Tevatron 
are still difficult to trigger on and identify.  
Excited electrons can decay via:
 $e^{*} \rightarrow e \gamma $, $e^{*}
\rightarrow e Z $, and $ e^{*}\rightarrow W \nu$ channels with the branching 
ratios shown in Fig.~\ref{BranchingRatios}.  The backgrounds to the $W$ and $Z$ 
decay channels in the hadronic channels are fairly large and the leptonic  branching fractions needed to
identify the $W$ and $Z$ channels are small; however, the backgrounds to the 
photonic final states are relatively small in comparison making them a 
gold-plated signature.

\begin{figure}[ht]
  \vspace*{-.75in}
  \epsfbox{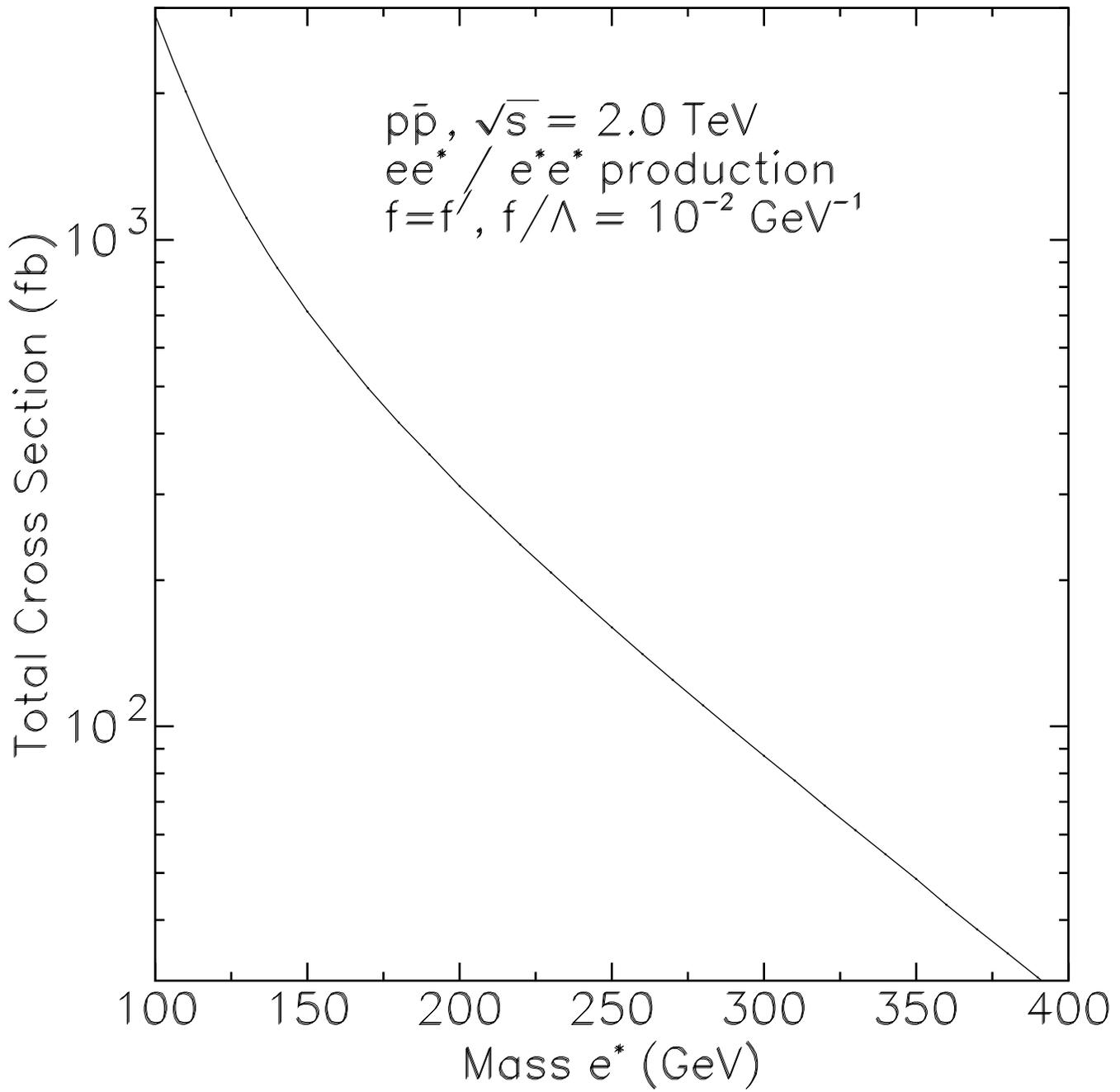}
  \vspace*{-1.25in}
  \caption{Total cross section for the sum of both single and pair 
  production and decay of excited electrons in the $ee\gamma$ final state.}
  \label{CrossSections}
  \end{figure}

In the case of the single excited electron production mode, there are two
possible signal 
$p \overline{p} \rightarrow  e^*e \rightarrow e^+ e^- \gamma $ and
$p \overline{p} \rightarrow e^*\nu \rightarrow e \nu \gamma$.
Similarly, pair production gives
$p \overline{p} \rightarrow  e^{*+}e^{*-} \rightarrow e^+ e^- \gamma\gamma $ 
and
 $p \overline{p} \rightarrow e^*\nu^* \rightarrow e \nu \gamma\gamma$.
For simplicity we concentrate on inclusive $ee\gamma$ final state for 
both single and pair production as it covers most of the production in both 
single and pair production cases.  Before taking into account the branching 
ratio to photons, the total production cross section is shown in 
Fig.~\ref{CrossSections} for the case $f/\Lambda = 10^{-2}~{\rm GeV}^{-1}$ and 
$f=f^{'}$.

There are a number of backgrounds to the $ee\gamma$ channel.  The dominant 
backgrounds are $W\gamma$+jets and $Z\gamma$+jets production.  Others include 
$W$+jets and $Z$+jets, and multijets, where jets can fake leptons and/or 
photons.  Studies have shown~\cite{DZero Diboson Studies} that the fake 
backgrounds can be modeled by the kinematics of the irreducible backgrounds.
The backgrounds are modeled using the same {\sc CompHEP} simulation structure 
described above.
Since there is an infrared singularity at $p_T^{\gamma} = 0$,
we require  $ P_{T}^{\gamma}> 10$ GeV and $\Delta R_{ij}>0.1 $, where 
$\Delta R_{ij}\equiv \sqrt{(\Delta\eta)^{2}+(\Delta\varphi)^{2}}$, and  
at the generator level, $i$ and $j$ are any lepton-photon combination.  
We take the K-factor, which has a value of about 1.36 depending on 
$p_{T}^{\gamma}$, for the background processes from the literature 
\cite{K-background}.

\begin{figure}[ht]
  \epsfbox{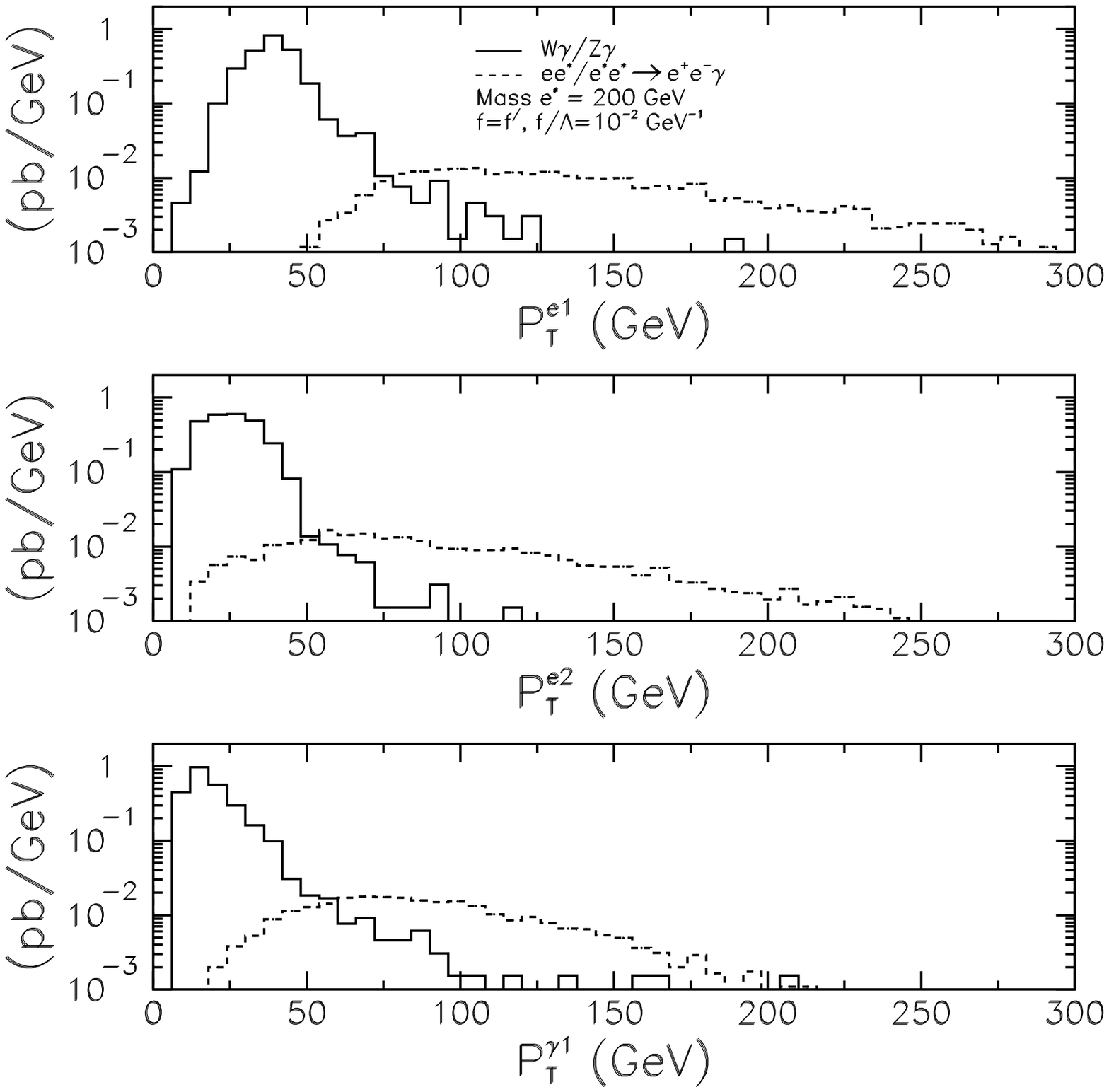}
  \vspace*{-1.25in}
  \caption{A comparison of the kinematic distributions of the final state 
  electrons and photons for excited electron and standard model background 
  processes.  The figures show the results for single and pair production of 
  ${e^*}e \rightarrow ee\gamma$ and ${e^*}{e^*} \rightarrow ee\gamma$.}
  \label{Kine1}
  \end{figure}

\begin{figure}[ht]
  \vspace*{-.75in}
  \epsfbox{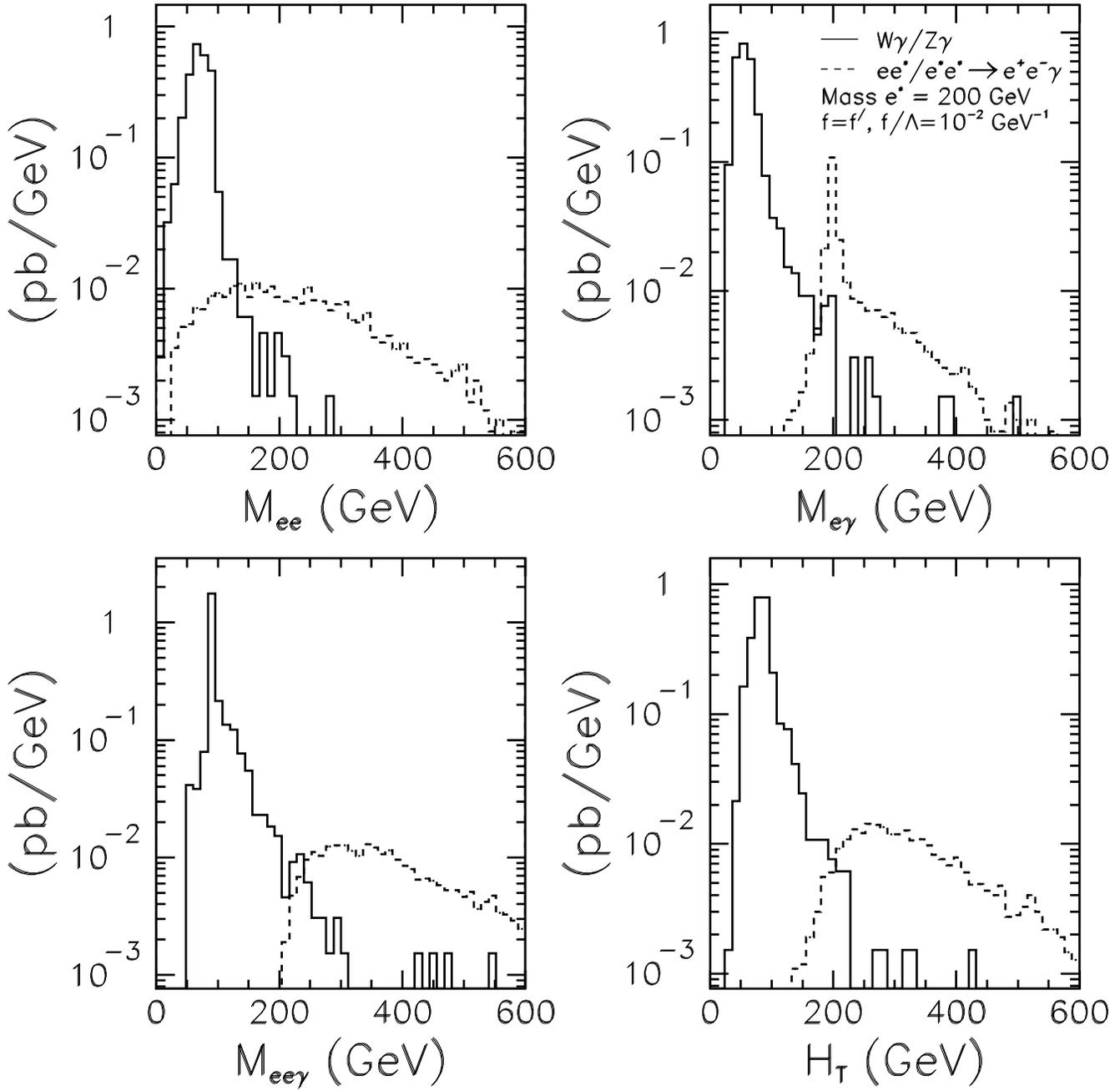}
  \vspace*{-1.25in}
  \caption{A comparison of some of the variables studied which gave good 
  separation between excited leptons and standard model background processes 
  in the $ee\gamma$ final state. Note that the $M_{e\gamma}$ variable gives 
  the best separation between signal and background for $M_{e^{*}}=200$ 
  GeV at $M_{e\gamma}>185$ GeV.}
  \label{Kine2}
  \end{figure}

For both signal and backgrounds, we use a parametric simulation to model
the detector response. The SHW detector simulation~\cite{SHW} has been 
shown to be an effective averaging between the CDF~\cite{CDF} and 
D\O~\cite{DZero} detectors for Run~II.  After detector simulation the kinematic
distributions for both the signal and estimated backgrounds are shown 
in Figs~\ref{Kine1} and~\ref{Kine2}.  We find acceptances at about 
the 0.3 level for signal.


To maximize our sensitivity we assume, for simplicity, that taking a set 
of cuts which minimizes the expected cross section limit also
maximizes our sensitivity to calculate these limits. To calculate these limits, we use the signal 
acceptance and background estimates from the simulations above. We add in a 
single detector with 2 fb$^{-1}$ of luminosity and we assume the experimental 
parameters of Table~\ref{Parameters} for systematic uncertainties.  We use a 
frequentist method~\cite{Frequentest} to incorporate the errors and assume 
that all errors on  acceptance, background and luminosity are uncorrelated.  
To determine the expected limit as a function of a given cut using Run~II data,
under the assumption of no observed signal, the 95\% confidence level (C.L.) cross 
section upper limit, for a given cut, is uniquely determined by number of events 
observed in the data.  With no signal, the number of events observed 
in the data, $N$, fluctuates around the mean number of expected events, $M$, 
according to Poisson statistics.  Thus, an average expected cross section 
limit is given by:

\begin{equation}
 \label{Bruce's Relation}
   \sigma_{95}^{cut}=\sum_{N=0}^{\infty} \sigma_{95}(N,cut)\frac{ e^{-M}M^{N}}{N!}
\end{equation}

\begin{table}[tb]
\centering
 \begin{tabular}{|c|c|}
  \hline
       Luminosity (fb$^{-1}$)&  2.0 $\pm$ 0.1\\
  \hline
        Background error  $\frac{\delta N_{back}}{N_{back}}$  &  0.1\\

  \hline
        Acceptance error  $\frac{\delta A}{A}$  &  0.1\\

  \hline
 \end{tabular}

 \caption{Parameters used to calculate the 95\% C.L.  cross section upper limits.}
 \label{Parameters}
\end{table}

\noindent In this way we get an average expected cross section limits as a function of each cut. We can then optimize for each mass
point as a function of the cuts.

After a study of multi-dimensional distributions for background and signal, we determined that the $e^+e^-\gamma$ final state has the best signal to background ratio and kinematical distributions. Furthermore, we find that applying a single cut on the kinematical variable $M_{e\gamma}$
 (where $e$ is the electron with highest $p_{T}$) gives the best relation
 between acceptance and $N$ to find minimal value of $\sigma_{95}$
 for a given mass of the excited electron. This is readily apparent
using Figs.~\ref{Kine1} and \ref{Kine2} which shows that the best separation
between signal and background is the $M_{e\gamma}$ variable.

\begin{table}[tb]
 \centering
 \begin{tabular}{|c|c|c|c|c|c|c|}
 \hline
 $M_{e^*}$ & $\sigma^{prod}$ & $\cal B$ ($e^{*} \rightarrow e\gamma$) &  Cut & N$_{back}$ & $\sigma_{\rm 95\%~C.L}^{expected}$ & $\sigma_{\rm 95\%~C.L}^{expected}\cdot\cal B$ \\
 (GeV) & (fb) &    & (GeV) & & (fb) & (fb)\\ \hline
  150 & 712 & 0.400 & 140.0  & 42.0 & 41.3 & 16.5\\
 \hline
  200 & 312  & 0.334  & 185.0 & 18.0 &  30.9 & 10.3\\
 \hline
  250 & 160  & 0.309 & 230.0 & 7.64 & 23.7 & 7.3\\
 \hline
  300 & 87 & 0.297  & 275.0 & 4.36 & 20.0 & 5.9\\
 \hline
\end{tabular}

\caption{Values for $\frac {f}{\Lambda}$ = $\frac {1}{100}$ GeV$^{-1}$ optimized points.}
\label{optimized}
\end{table}

\begin{figure}[ht]
  \vspace*{-.75in}
  \epsfbox{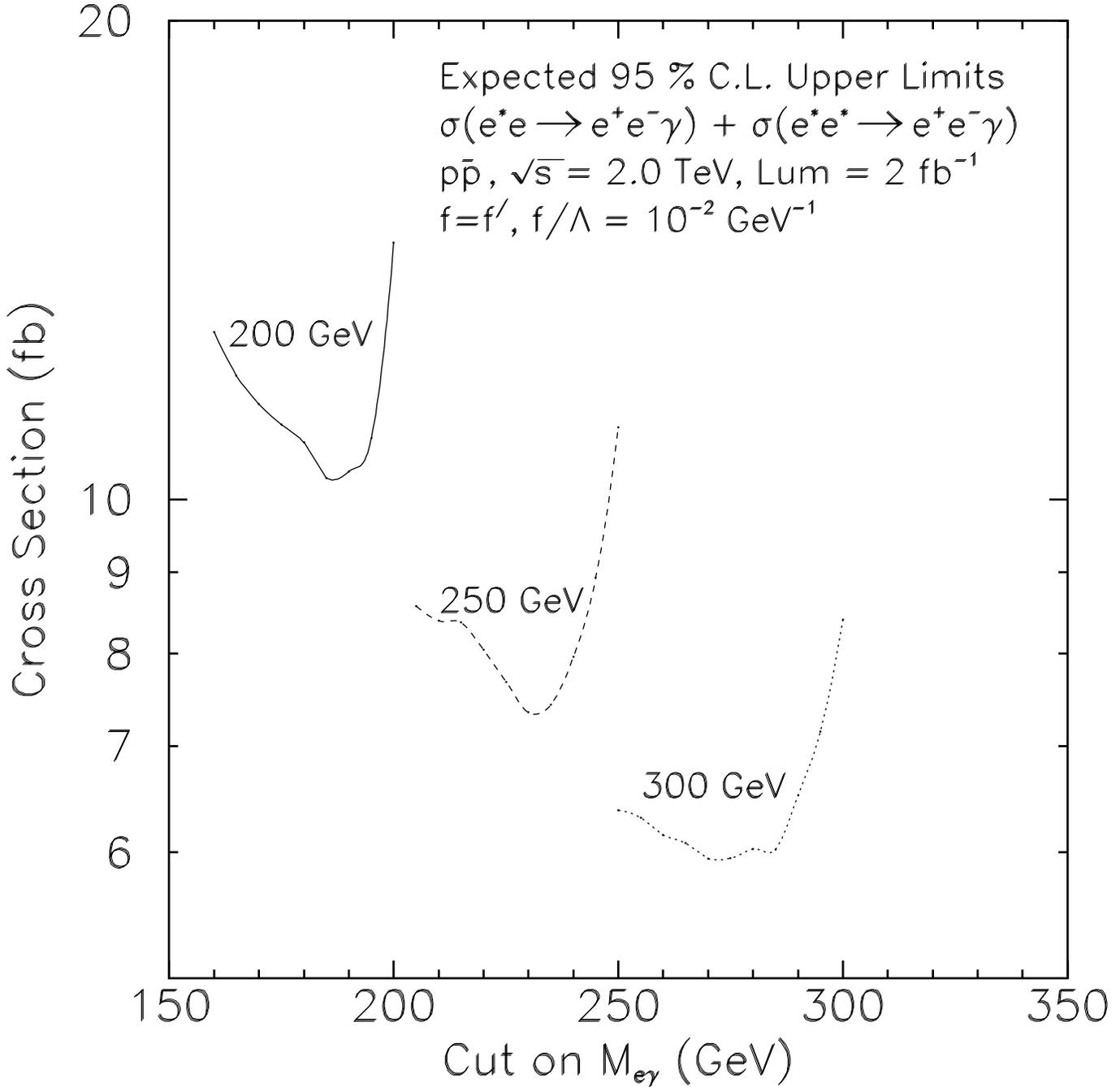}
  \vspace*{-1.25in}
  \caption{The expected 95\% C.L. cross section upper limit as a function 
  of the $M_{e\gamma}$ cut for different masses of $e^{*}$.  There the cross 
  section is defined as $\sigma (p\bar{p} \rightarrow ee^{*} \rightarrow 
  ee\gamma)$ + $\sigma(p\bar{p} \rightarrow e^{*}e^{*} \rightarrow 
  ee\gamma)$ and $\mathcal {L}$ = 2 fb$^{-1}$.}
  \label{sigma95}
  \end{figure}

\begin{figure}[ht]
  \vspace*{-1.75in}
  \epsfbox{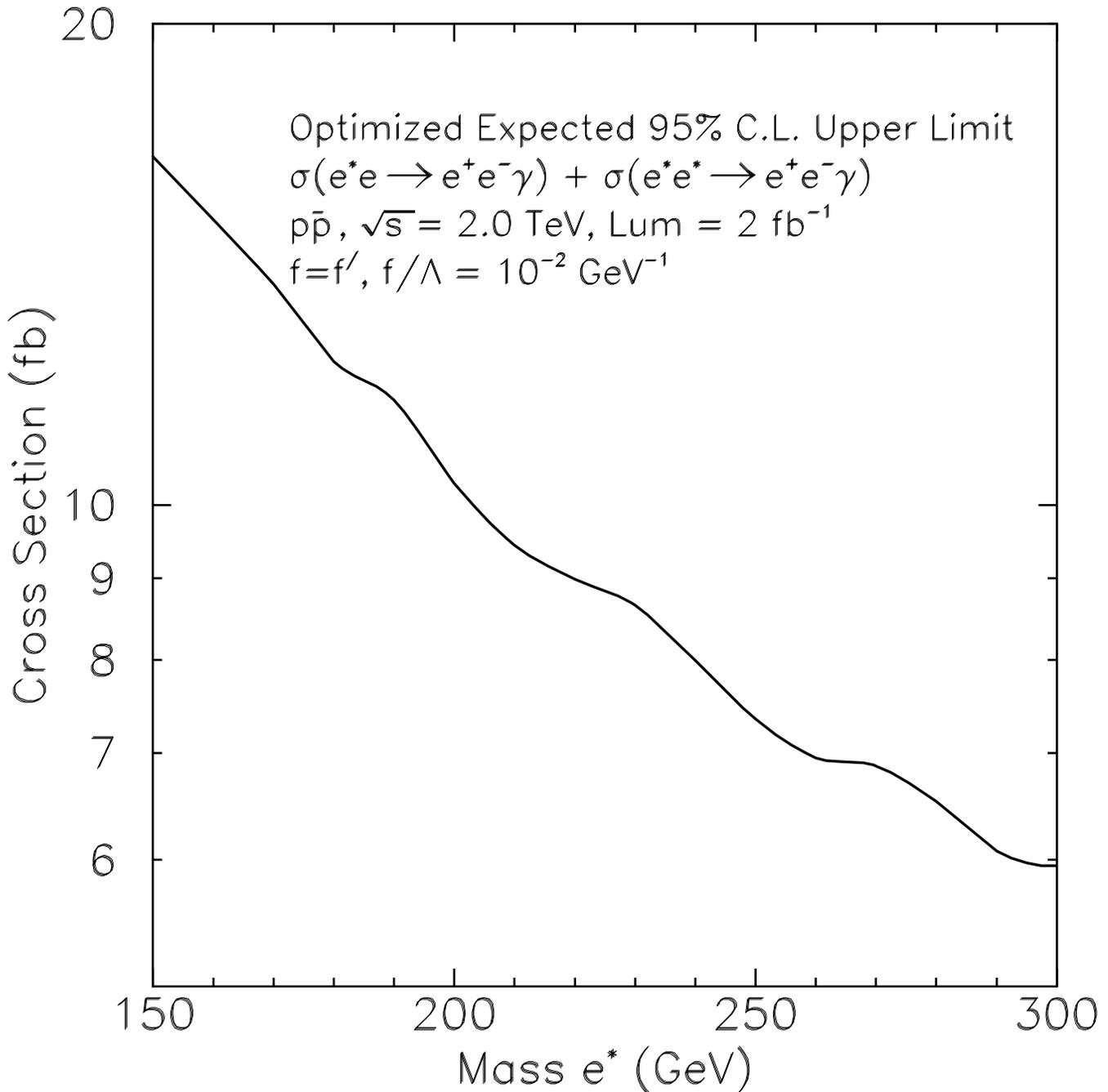}
  \vspace*{-1.25in}
  \caption{Optimized expected 95\% C.L. for $e^* \rightarrow ee\gamma$ 
  production using the $ee\gamma$ final state and 2 fb$^{-1}$ of data with a 
single detector.}
  \label{cross95}
  \end{figure}

Figure~\ref{sigma95} shows the expected 95\% C.L. cross
section upper limit, $\sigma_{95}$, as a function of the $M_{e\gamma}$ cut for different masses of $e^{*}$ for $\frac{f}{\Lambda}$ = $\frac{1}{100}$ GeV$^{-1}$. Placing our cut at the minimum of the curve gives our optimization, and final expected cross section limit. These results are shown in Fig.~\ref{cross95}. Using the
 Feynman rules for the Lagrangian in Eqn.~\ref{EffLagrangian} leads to a simple relation  between cross-section and $\frac{f}{\Lambda}$ for signal:

\begin{eqnarray}
(\Lambda_{1})^{2}\sigma(\frac{f}{\Lambda_{1}},M_{e^{*}})=(\Lambda_{2})^2\sigma(\frac{f}{\Lambda_{2}},M_{e^{*}}) \\
\frac{f}{\Lambda} = \frac{ \sigma_{95} }{10^{4}\sigma(10^{-2},M_{  e^{*} })}
\end{eqnarray}\\

\begin{figure}[ht]
  \vspace*{-.75in}
  \epsfbox{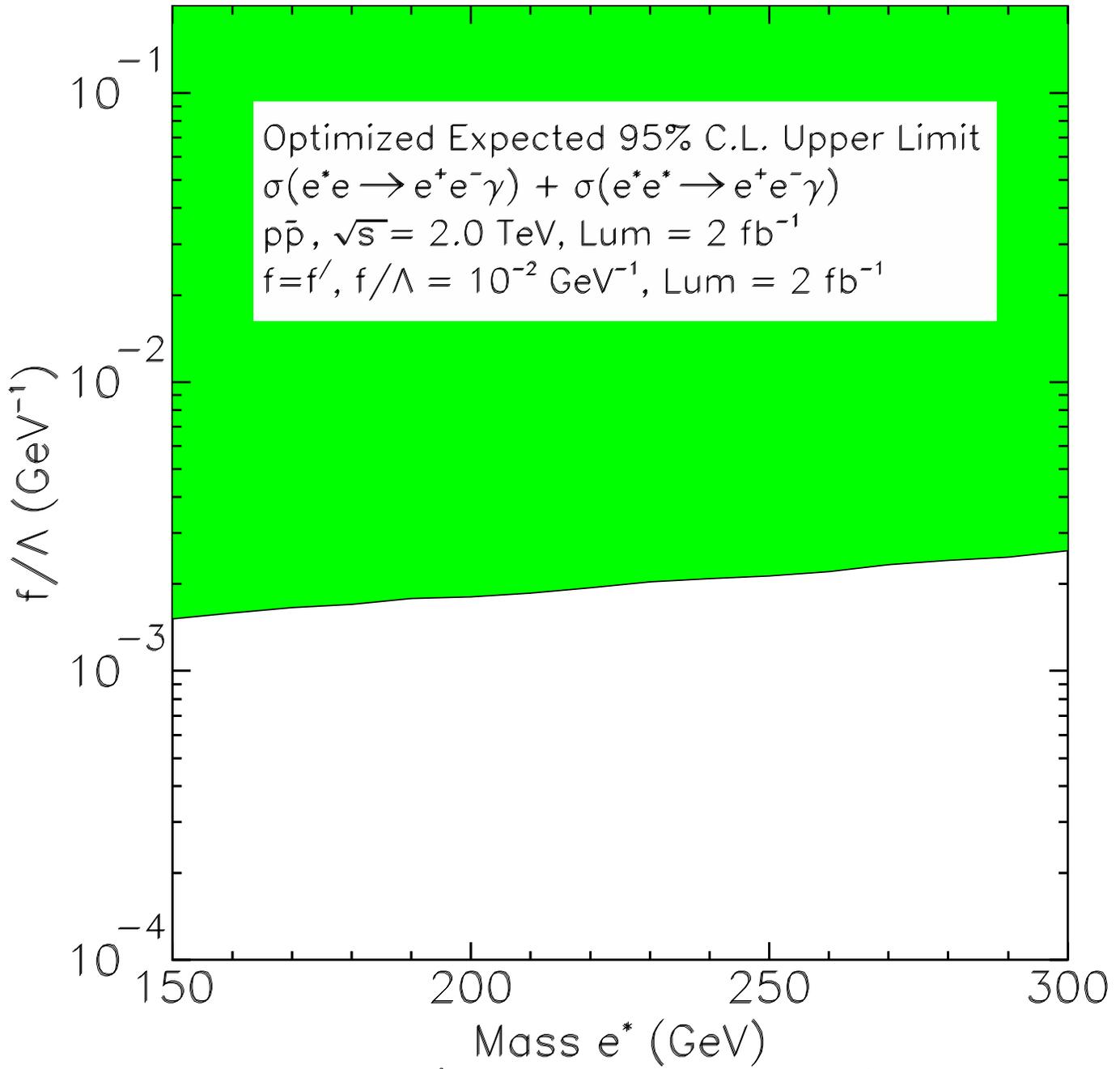}
  \vspace*{-1.25in}
  \caption{Exclusion plot for $\frac{f}{\Lambda}$ as a function of $e^*$ mass
  using 2 fb$^{-1}$ worth of data and a single detector.}
  \label{ExclPlot}
  \end{figure}

\begin{figure}[ht]
  \vspace*{-1.75in}
  \epsfbox{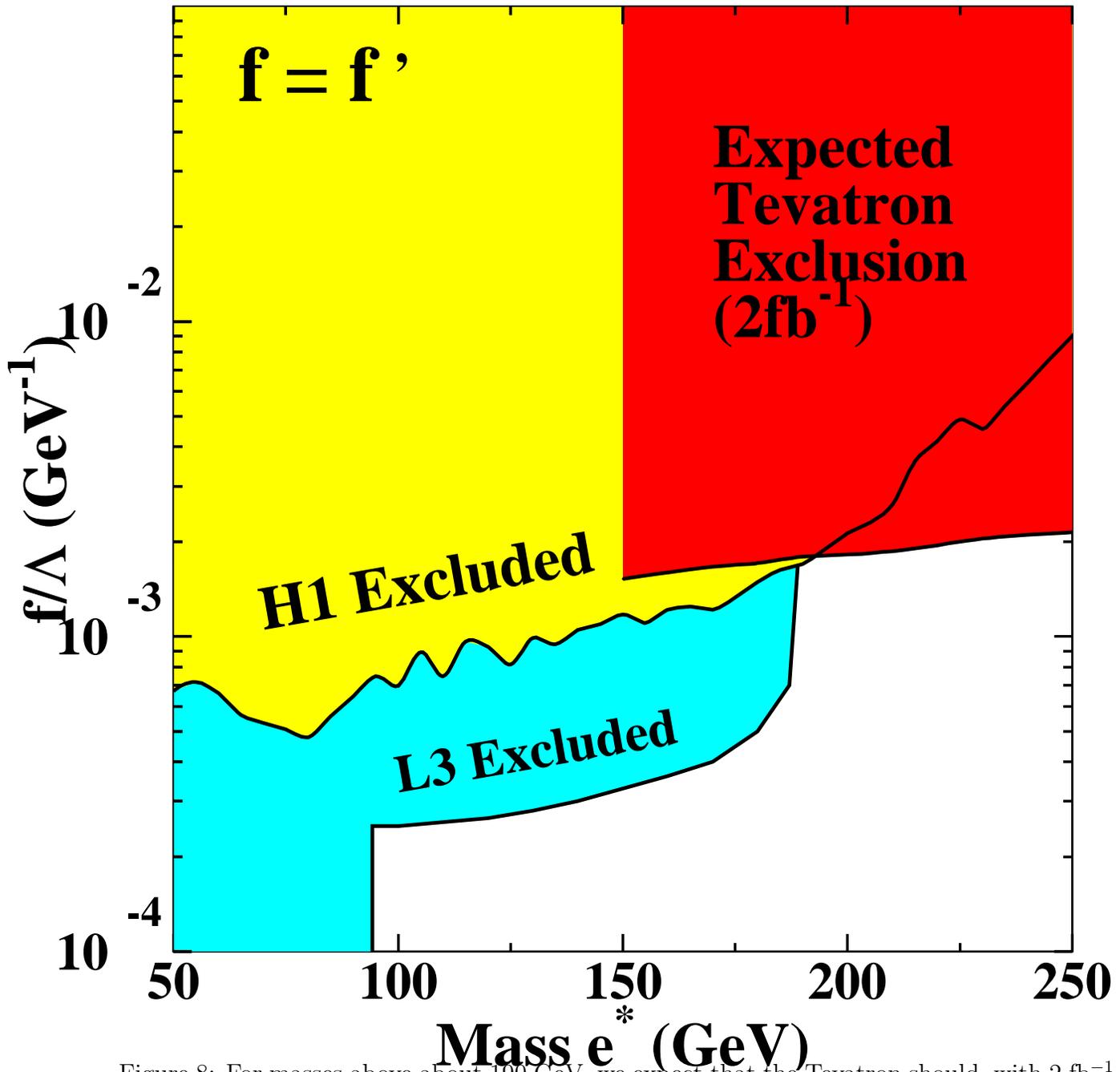}
  \vspace*{-1.25in}
  \caption{For masses above about 190 GeV, we expect that the Tevatron should, with 2 fb$^{-1}$ of data, and one detector, produce the most stringent limits.}
  \label{Result}
  \end{figure}

Taking into account this equation, numerical values of signal
cross-sections (where $\frac{f}{\Lambda} = 10^{-2}$ GeV$^{-1}$)
 and results for $\sigma_{95}$ we can make an  exclusion plot in the 
$\frac{f}{\Lambda}$ vs. $M_{e^{*}}$ plane. This result is shown in 
Fig.~\ref{ExclPlot}.  We compare our results to those of LEP and HERA
in Fig.~\ref{Result} which would give the most stringent limits, to date, for 
masses above 190 GeV.

The prospects for searching for excited electrons
at the upgraded Fermilab Tevatron are excellent. We expect that with a single detector and 2 fb$^{-1}$ of data should significantly extend the mass
reach, especially in the low $\frac{f}{\Lambda}$ region and large masses.
We also point out that similar results are obtained for excited muons
which significantly improve on the current limits which are not producible at HERA.

The authors would like to thank John Hobbs for the use of his limit calculator, Bruce Knuteson for his relation in Eqn.~\ref{Bruce's Relation}, and Bhaskar Dutta for helpful discussions.  We would also like to thank the D$\O$
collaboration for the use of their computers to do the simulation work.
We also would like to thanks the Institute of Nuclear Physics of Moscow State University, the University of Maryland and Texas A\&M University for their support.  Finally we would like to thank the US DOE and Russian Ministry of Industry, Science and Technology for their support during this project. The work of E.B. and A.V. was partly supported by the RFBR-DFG 99-02-04011, RFBR 00-01-00704, CERN-INTAS 99-377, and Universities of Russia
990588 grants.

\clearpage









\clearpage
 
\clearpage



\end{document}